\documentclass[pra,twocolumn,floatfix,superscriptaddress,longbibliography,notitlepage]{revtex4-2}

\usepackage{graphicx}
\usepackage{dcolumn}
\usepackage{bm}

\usepackage{subfig}
\usepackage{amsthm}
\usepackage{amssymb}
\usepackage{amsmath}
\usepackage{verbatim}
\usepackage{dcolumn}
\usepackage{bm}
\usepackage{epsf}
\usepackage{color}
\usepackage[colorlinks=true,citecolor=blue,linkcolor=blue,urlcolor=blue]{hyperref}%
\usepackage{xcolor}
\usepackage{dsfont}

\begin{document}


\title{Efficiency Optimization in Quantum Computing: \\Balancing Thermodynamics and Computational Performance}

\author{Tomasz Śmierzchalski}
\email{tsmierzchalski@iitis.pl}
\author{Zakaria Mzaouali}%
 \email{zmzaouali@iitis.pl}
\affiliation{%
Institute of Theoretical and Applied Informatics, Polish Academy of Sciences, Bałtycka 5, Gliwice, 44-100, Poland.
}
\author{Sebastian Deffner}
\email{deffner@umbc.edu}
\affiliation{Department of Physics, University of Maryland, Baltimore County, Baltimore, MD 21250, USA}

\author{Bartłomiej Gardas}
 \email{bgardas@iitis.pl}
\affiliation{%
Institute of Theoretical and Applied Informatics, Polish Academy of Sciences, Bałtycka 5, Gliwice, 44-100, Poland.
}

\date{\today}

\begin{abstract}
We investigate the computational efficiency and thermodynamic cost of the D-Wave quantum annealer under reverse-annealing with and without pausing. Our experimental results demonstrate that the combination of reverse-annealing and pausing leads to improved computational efficiency while minimizing the thermodynamic cost compared to reverse-annealing alone. Moreover, we find that the magnetic field has a positive impact on the performance of the quantum annealer during reverse-annealing but becomes detrimental when pausing is involved. Our results provide strategies for optimizing the performance and energy consumption of quantum annealing systems employing reverse-annealing protocols.
\end{abstract}

\maketitle

\section{\label{intro} Introduction}
Large scale investments in quantum technologies are usually justified with promised advantages in sensing, communication, and computing \cite{Raymer2019QST}. Among these, quantum computing is probably the most prominent application, since it has the potential to revolutionize information processing and computational capabilities. 
For certain tasks, quantum computers exploit the fundamental principles of quantum mechanics to perform complex calculations exponentially faster than classical computers~\cite{mcmahon2007quantum,nielsen2002quantum,Sanders2017}. The tremendous computational power offered by quantum systems has fueled excitement and exploration in various scientific, industrial, and financial sectors~\cite{Raymer2019QST, quantum_EU, quantum_Japan, quantum_Canada, quantum_UK, quantum_Australia, health, finance, 2023_Domino}. Consequently, there have been significant developments in the pursuit of quantum advantage that have propelled quantum computing from theoretical speculation to practical implementation~\cite{silicon1998, 5qubit_2000, 12qubit_2006, shor_algo_2007, deutsch_algo_2007, dwave2011, QC_RT_2013, QS_2019, QD_2019, QD_cho_2019, entanglement_2023}. 

Major technology companies such as IBM, Google, Microsoft, Intel, and Nvidia have been investing massively in quantum research and development, leading to the establishment of quantum computing platforms and open-source frameworks that enable researchers and developers to experiment and explore the potential of quantum algorithms and applications~\cite{companies_QC}. These advancements have been driven by breakthroughs in both hardware and algorithmic techniques, bringing us closer to realizing the potential of quantum computers~\cite{survey1, survey2}.

However, the rapid development of quantum technologies also raises critical questions about the energy requirements and environmental implications of quantum computation~\cite{deffner_campbell,auffeves}. Energy consumption has become a focal point for researchers, policymakers, and society at large, as the demand for computing power continues to rise, and concerns about climate change and sustainability intensify. Consequently, assessing the energy consumption of quantum computers is vital for evaluating their feasibility, scalability, and identifying potential bottlenecks~\cite{review_energy_QC}.

The energy consumption of quantum computers stems from various sources, including the cooling systems needed to maintain the delicate quantum states, the control and manipulation of qubits, and the complex infrastructure required to support quantum operations~\cite{2017_energy_QC, fellous2022optimizing}. The superposition characteristic of qubits demands a sophisticated physical environment with precise temperature control and isolation, leading to significant energy expenditures. These challenges call for synergic work between quantum information science, quantum engineering, and quantum physics, to develop an interdisciplinary approach to tackle this problem~\cite{likharev1982classical}.

The theoretical framework to quantify the energy consumption of quantum computation is through quantum thermodynamics, which provides the necessary tools to quantify and characterize the efficiency of emerging quantum technologies and therefore is crucial in laying a roadmap to scalable devices~\cite{gemmer2009quantum, QT_anders, deffner_campbell}. Quantum thermodynamics assess the thermodynamic resources required to process and manipulate quantum information~\cite{goold2016role, binder2018thermodynamics,deffner_campbell}, with a notable focus on the fundamental limits such as quantum versions of Landauer's principle~\cite{Gea2002PRL,Bedingham2016NJP,Cimini2020,Timpanaro2020PRL,Deffner2021EPL}. The exploration of the thermodynamics of information is not limited to the equilibrium settings, as recent research has delved into the nonequilibrium aspects of quantum computation, particularly in the context of quantum algorithms and quantum simulation~\cite{nonequilibrium_QC_2018, nonequilibrium_QC_2022}. Understanding the thermodynamics of quantum systems, including the generation of entropy, heat dissipation, and non-equilibrium dynamics, serves into optimizing the algorithmic performance, energy consumption, and resource utilization~\cite{meier2023energyconsumption}. 

The study of the thermodynamics of quantum computers has been an active research area with notable results that deepen our understanding of the energy landscapes, heat dissipation, and efficiency of quantum computation while also addressing challenges related to noise, decoherence, and thermal effects~\cite{2015_Gardas,2016_Gardas,2016_Gardas_PRA,2018_Gardas, 2018_Gardas_Nature}.  The optimization of the energy efficiency of quantum computers has been approached from several angles, for instance to eluciate the minimization of the energy dissipation during computations, and to develop energy-efficient algorithms and architectures~\cite{2021_Mzaouali_PRE, 2022_Soriani_PRA, 2022_Soriani_PhysRevA, 2022_Coopmans_PRR, 2022_Kazhybekova_IOP, 2023_Xuereb_PRA, carolan_2023, 2023_Kiely_PRR, 2023_Zawadzki_PRA}. The focus is on reducing energy requirements and increasing the computational efficiency of quantum systems, paving the way for sustainable and practical quantum computing technologies. As quantum computers generate heat during operation, effective thermal management becomes essential to maintain qubit stability and mitigate thermal errors. 

In this paper, we study the interplay between thermodynamic and computational efficiency in the quantum annealing. In recent years, thermodynamic considerations of the D-Wave quantum annealer have become prevalent. For instance, some of us used the quantum fluctuation theorem to assess the performance of annealing~\cite{2018_Gardas_Nature}. Furthermore, the working mechanism of the D-Wave chip was shown to be equivalent to that of a quantum thermal machine, e.g. thermal accelerator, under the reverse-annealing schedule~\cite{buffoni2020thermodynamics}. Here, we take a step further and analyze the energetic and computational performance of quantum annealing under reverse-annealing, and how to optimize it through the introduction of pausing in the annealing schedule. We perform our experiments on the D-Wave $2000Q$ quantum annealer and we show that a pause in the annealing schedule allows us to achieve better computational performance at a lower energetic cost. Additionally, we discuss the role and impact of the magnetic field on the performance of the chip. 


\section{\label{theory} Theory \& Figures of Merit}

\begin{figure}
    \includegraphics[width=.48\textwidth]{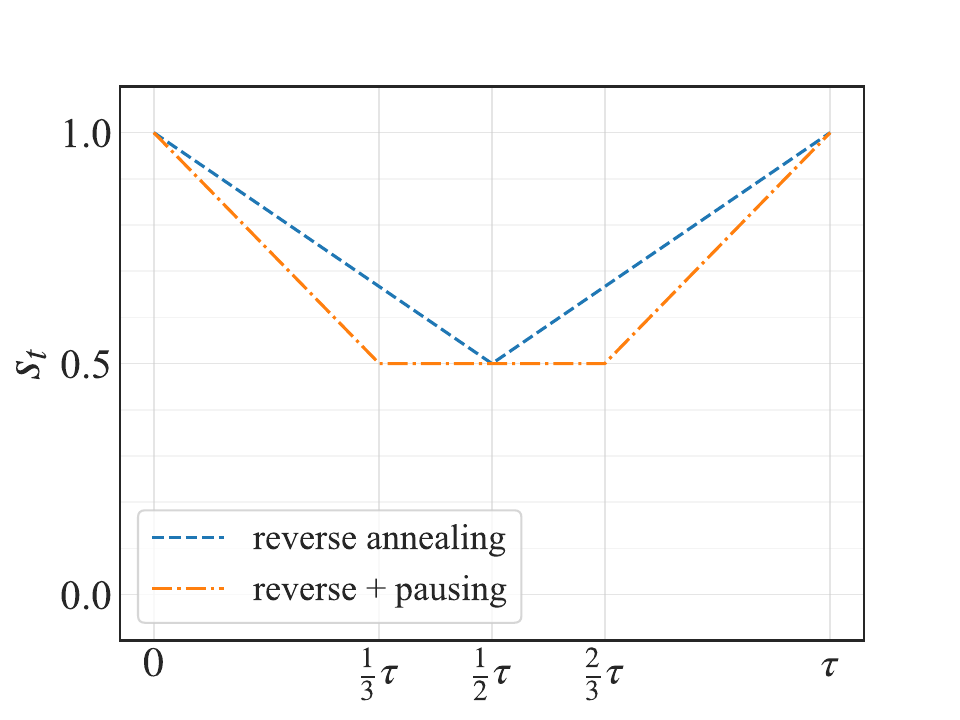}
    \caption{Quantum annealing protocols used in experiments. $s_t$ is an annealing parameter, $s_t = t/\tau$, $t$ represent time and $\tau$ is the total annealing time in $\mu$s.}
    \label{QA_protocols}
\end{figure}

We start by briefly outlining notions and notations. Quantum annealing consists of mapping the optimization problem to a mathematical model that can be described using qubits~\cite{PRE_Nishimori}. The quantum annealer is initialized in a quantum state that is easy to prepare. The system is then evolved according to a time-varying Hamiltonian, which is a mathematical operator representing the problem's objective function and can be expressed as,
\begin{equation}
    H(s_t)=(1-s_t)\sum_i \sigma_i^x + s_t \left( \sum_i h_i \sigma_i^z + \sum_{\langle i,j \rangle} J_{i,j} \sigma_i^z \sigma_j^z \right),
    \label{ham}
\end{equation}
where $\sigma_i^{\alpha}$, with $\alpha=(x,z)$ are Pauli matrices, and $h_i$ is the local magnetic field. $s_t=t/\tau$ describes the annealing schedule which controls the rate of the transformation between the easy-to-prepare Hamiltonian $H_0=\sum_i \sigma_i^x $ and the problem specific Hamiltonian $H_p= \sum_i h_i \sigma_i^z + \sum_{\langle i,j \rangle} J_{i,j} \sigma_i^z \sigma_j^z $, with $\tau \in [0,t]$. On the D-Wave machine, the annealing time $t$ can be chosen from microseconds ($\sim\!2\mu s$) to milliseconds ($\sim\!2000 \mu s$).

The usual quantum annealing process, called forward annealing, starts with initializing the qubits in a known eigenstate of $\sigma^x$. The system is then slowly driven by varying the Hamiltonian parameters~\cite{albash2018adiabatic}. Initially, the driver Hamiltonian $H_0$ dominates, and the qubits are in a quantum superposition. As the annealing progresses, the problem Hamiltonian $H_p$ gradually becomes dominant, and the qubits tend to settle into the low-energy states that correspond to the optimal solution of the problem. 

In this work, we employ a different protocol called reverse-annealing as depicted in Fig.~\ref{QA_protocols}, where the processor initially starts with a classical solution defined by the user to explore the local space around a known solution to find a better one. Reverse-annealing has been shown to be more effective than forward annealing in some specific use cases, including nonnegative/binary matrix factorization~\cite{golden2021reverse}, portfolio optimization problems~\cite{venturelli2019reverse}, and industrial applications~\cite{yarkoni2022quantum}. Moreover, reverse-annealing has unique thermodynamic characteristics with typically enhanced dissipation~\cite{buffoni2020thermodynamics,buffoni2021thermodynamics}.

\begin{figure}
    \includegraphics[width=.48\textwidth]{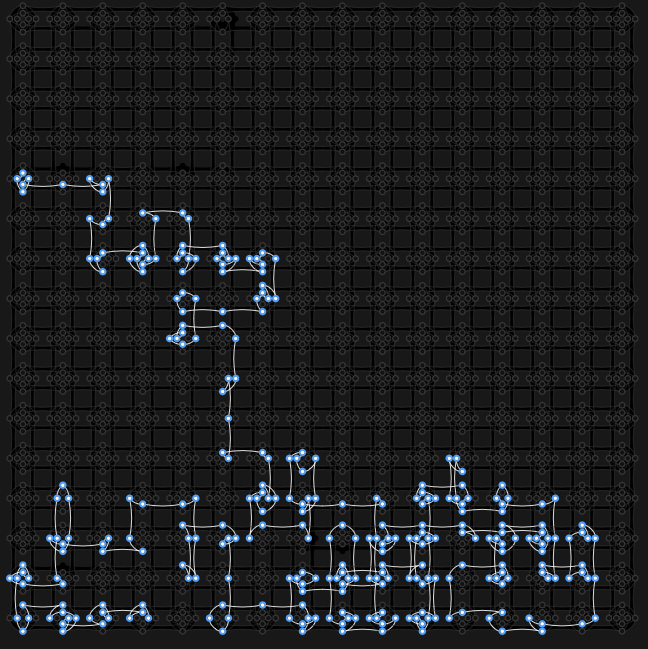}
    \caption{Example of the embedding of the $300-$spin Ising chain onto D-Wave $2000Q$ quantum processing unit (QPU) with the chimera architecture~\cite{pegasusChimera}. White-blue dots and lines are active qubits, and grey ones are inactive. This is one-to-one embedding, such that every physical qubit corresponds to one logical qubit.}
    \label{connectivity}
\end{figure}

To quantify the thermodynamic efficiency of the D-Wave $2000Q$ chip, we initialize the quantum annealer in the spin configuration described by a thermal state at inverse temperature $\beta_1=1$, and we assume that initially the system+environment state is given by the joint density matrix,
\begin{equation}
    \rho=\frac{\exp{(-\beta_1 H_S)}}{Z_S} \otimes \frac{\exp{(-\beta_2 H_E)}}{Z_E}.
\end{equation}
The energy transfer between two quantum systems initially at different temperatures is described by the quantum exchange fluctuation theorem~\cite{Jarzynski2004PRL,Sone2023AVSQS},
\begin{equation}
    \frac{p(\Delta E_1,\Delta E_2)}{p(-\Delta E_1,-\Delta E_2)}=e^{\beta_1 \Delta E_1 + \beta_2 \Delta E_2},
    \label{eq3}
\end{equation}
where $\Delta E_i, i=1,2$ are, respectively, the energy changes of the processor and its environment during the annealing time $t$, and $p(\Delta E_1,\Delta E_2)$ is the joint probability of observing them in a single run of the annealing schedule. Equation~\eqref{eq3} can be re-written in terms on the entropy production $\Sigma =\beta_1 \Delta E_1 + \beta_2 \Delta E_2$ as~\cite{Deffner2011PRL,Touil2021PRXQ},
\begin{equation}
    \frac{p(\Sigma,\Delta E_1)}{p(-\Sigma,-\Delta E_1)}=e^{\Sigma}. 
\end{equation}

Note that during the experiment, we have \emph{only} access to the energy change of the processor $\Delta E_1$. Therefore, the thermodynamic quantities  entropy production $\langle \Sigma \rangle$, average work $\langle W\rangle$, and average heat $\langle Q\rangle$ are not directly accessible. However, they can be lower bounded by thermodynamic uncertainty relations~\cite{TUR},
\begin{align}
    \langle \Sigma \rangle &\geq  2g\left(\frac{\langle \Delta E_1 \rangle}{\sqrt{\langle \Delta E_1^2 \rangle}}\right),\\
    -\langle Q \rangle &\geq \frac{2}{\beta_2}g\left(\frac{\langle \Delta E_1 \rangle}{\sqrt{\langle \Delta E_1^2 \rangle}}\right) - \frac{\beta_1}{\beta_2} \langle \Delta E_1 \rangle,\\
    \langle W \rangle &\geq \frac{2}{\beta_2}g\left(\frac{\langle \Delta E_1 \rangle}{\sqrt{\langle \Delta E_1^2 \rangle}}\right) + \left( 1 - \frac{\beta_1}{\beta_2} \right) \langle \Delta E_1 \rangle,
\end{align}
where $g(x)=x\tanh^{-1}{(x)}$, and $\beta_2$ is the temperature of the environment which can be estimated experimentally using the pseudo-likelihood method introduced in~\cite{beta2}. Accordingly, the upper bound on the thermodynamic efficiency can be determined from 
\begin{equation}
    \eta_{\text{th}} \leq - \frac{ \langle W \rangle}{ \langle Q \rangle}.
    \label{nth_eq}
\end{equation}
Moreover, we are interested in analyzing the computational efficiency of the quantum annealer, which we define as
\begin{equation}
    \eta_{\text{comp}} \leq  \frac{ \mathcal{P}_\text{GS}}{ \langle W \rangle},
    \label{ncomp_eq}
\end{equation}
and where have introduced the probability that ground state $s^\star$ is found in the given annealing run,
\begin{equation}
    \mathcal{P}_{\text{GS}} = \mathbb{P}(s^\star \in \text{\textbf{s}}).
    \label{pgs}
\end{equation}
This quantity is computed by dividing the number of successful runs (i.e., those which have found the ground state) by the total number of runs. The efficiencies defined by Eq.~\eqref{nth_eq} and Eq.~\eqref{ncomp_eq} are the main figures of merit of our analysis.

It is also instructive to analyze the ratio of the theoretical ($E_{\text{th}}=-299$) ground state energy and the experimental ($E_{\text{exp}}$) value read from the machine,
\begin{equation}
    \mathcal{F}_{\text{GS}} = \Big\langle \frac{E_{\text{exp}}}{E_{th}} \Big\rangle,
    \label{fgs}
\end{equation}
which is averaged over the number of samples.

\begin{figure*}
    \centering
    \subfloat[Success probability, Eq.~\eqref{pgs}, and fidelity, Eq.~\eqref{fgs}.]{\includegraphics[width=0.25\textwidth]{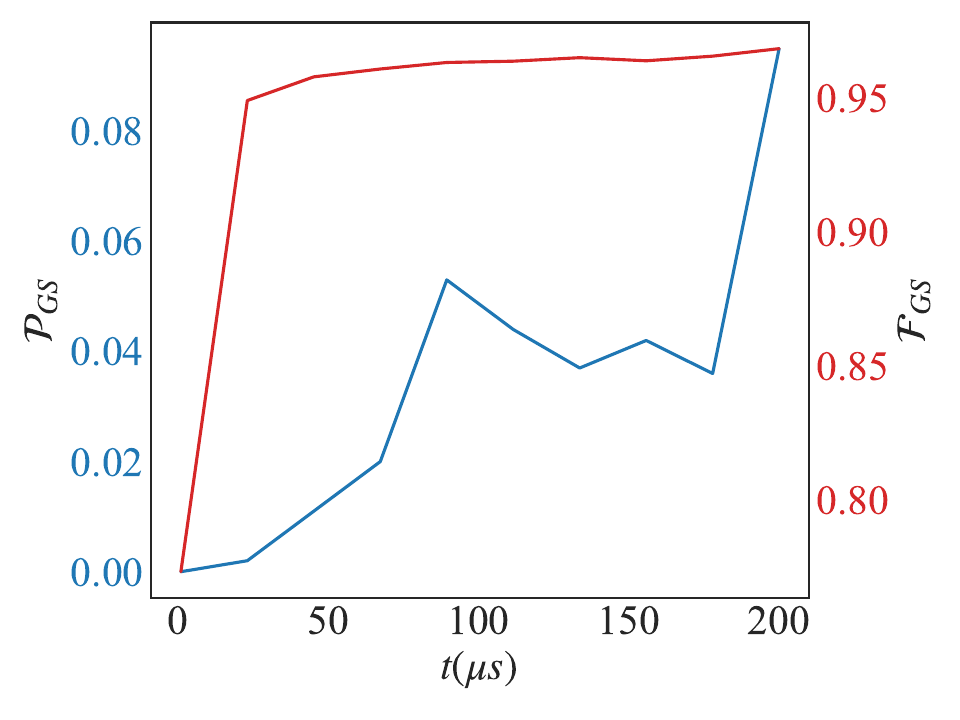}}%
    \subfloat[Thermodynamic and computational efficiency, Eq.~\eqref{nth_eq} and Eq.~\eqref{ncomp_eq} respectively.]{\includegraphics[width=0.25\textwidth]{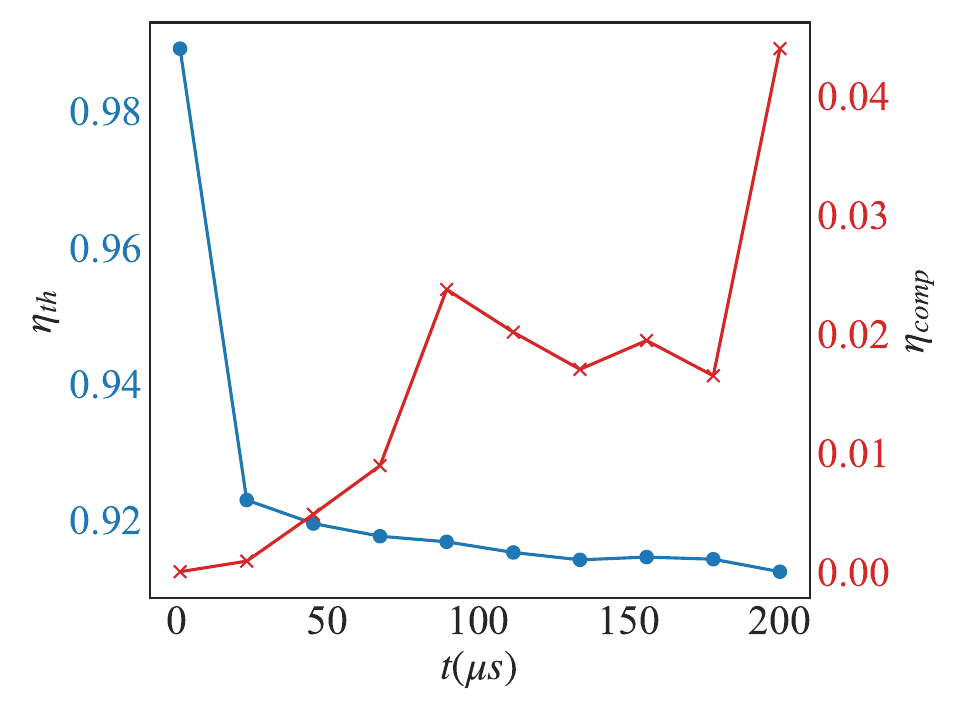}}
    \subfloat[Success probability, Eq.~\eqref{pgs}, and fidelity, Eq.~\eqref{fgs}.]{\includegraphics[width=0.25\textwidth]{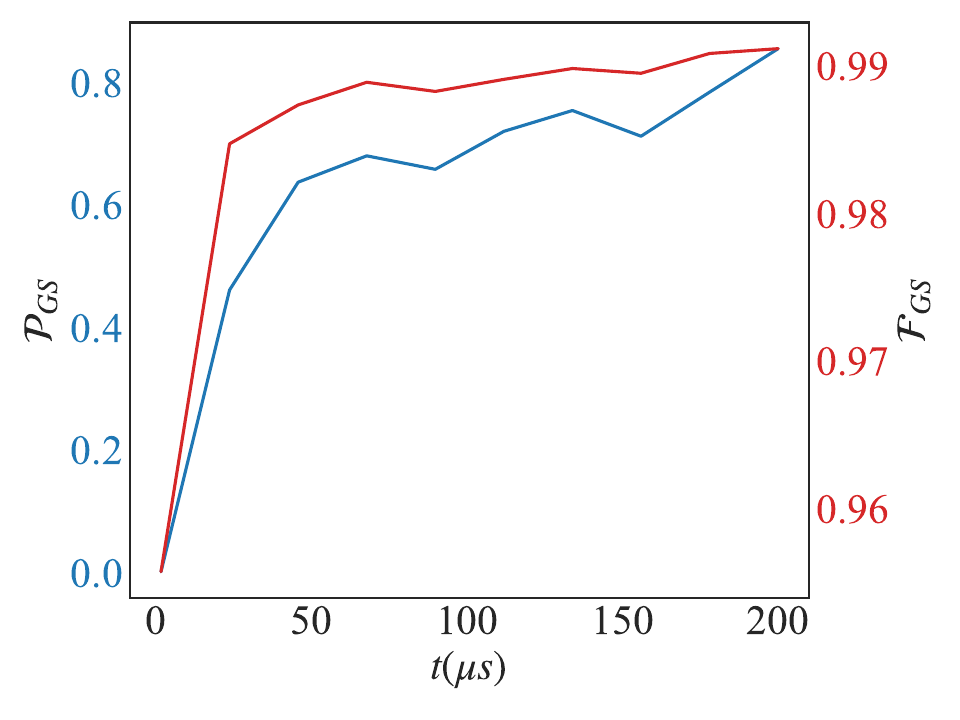}}%
    \subfloat[Thermodynamic and computational efficiency, Eq.~\eqref{nth_eq} and Eq.~\eqref{ncomp_eq} respectively.]{\includegraphics[width=0.25\textwidth]{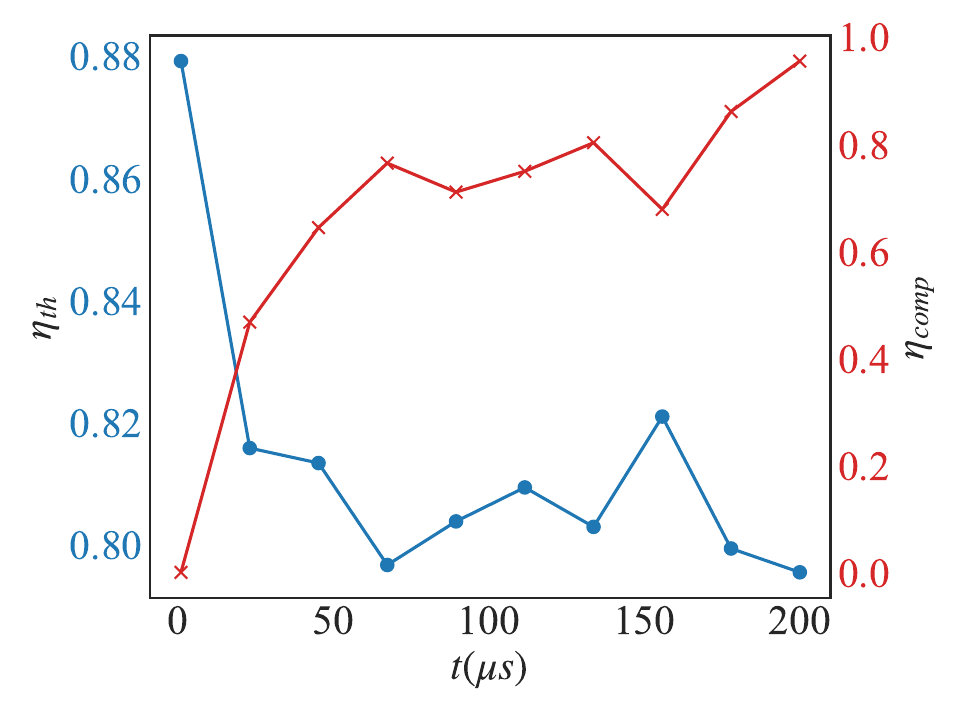}}
    \caption{Figures of merit under reverse-annealing. (a+b) without the pause and (c+d) with pause for $N=300$ spins, ferromagnetic couplings $J$, and zero magnetic field. Each point is averaged over 1000 annealing runs with 10 samples each.}
    \label{reverse_no_h}
    
\end{figure*}

\section{\label{Experiments} Experiments}

 All our experiments were performed on a D-Wave $2000Q$ quantum annealer. We considered an antiferromagnetic (i.e. $\forall i \, J_{i, i+1} = 1$) Ising chain on $N\!=\!300$ spins, with Hamiltonian as defined in~\eqref{ham}. However, one must first embed the given problem into the target quantum processing unit (QPU) architecture to perform quantum annealing. Here, embedding means finding a mapping between physical qubits presented in the machine and logical qubits (i.e. $\sigma^z_i$) representing our problem. Figure~\eqref{connectivity} shows an example embedding of our Ising problem onto the QPU with Chimera architecture.

We used annealing schedules shown in Fig.~\eqref{QA_protocols}. The system was initialized for both schedules by taking a sample thermal state at $\beta=1$ using Gibbs sampling \cite{mcmc}. For reverse annealing, with a given annealing time $\tau$, we ran $s_t \rightarrow 1/2$ for $t \in (0, \tau/2)$  and  $s_t \rightarrow 1$ for the remaining time. In reverse annealing with pausing $s_t \rightarrow 1/2$ for $t \in (0, \tau/3)$, next for $t \in (\tau/3, 2\tau/3)$ $s_t = 1/2$, as we pause the annealing process. Lastly, for $t \in (2\tau/3, 1)$ we let $s_t \rightarrow 1$.

\subsection{Zero magnetic field -- ``naked performance''}

We start with the case where the magnetic field $h$ is turned off. 

\paragraph*{Reverse-annealing without pausing.}

Figures~\eqref{reverse_no_h}(a+b) reports data from reverse-annealing experiments without the introduction of the pause. The success probability, Eq.~\eqref{pgs}, is shown in panel~(a) where we see that under reverse-annealing without the pause, the success probability of finding the ground state energy is very low and does not exceed 10\% in the whole annealing time. This means that in all 10000 samples taken in the experiment, only less than 1000 samples provided the ground state energy. However, although the probability to reach the ground state energy is low, its fidelity, Eq.~\eqref{fgs}, which is shown in the same panel is very high and saturates at$~0.95$.

The results shown in panels (a) are consistent with the thermodynamic and computational efficiency, Eq.~\eqref{nth_eq} and Eq.~\eqref{ncomp_eq} respectively, reported in panel (b). We see that under reverse-annealing without the pause, the computational efficiency of the chip $\eta_{comp}$ grows with the annealing time $t$ while being very low and not exceeding 4\% in the whole annealing time, which follows the behavior of the success probability $\mathcal{P}_{GS}$. Energetically, the thermodynamic efficiency $\eta_{th}$ decays exponentially with $\tau$ and remain at a high value close to $1$.
\begin{figure*}
    \centering
    \subfloat[$h=0.1$]{\includegraphics[width=0.33\textwidth]{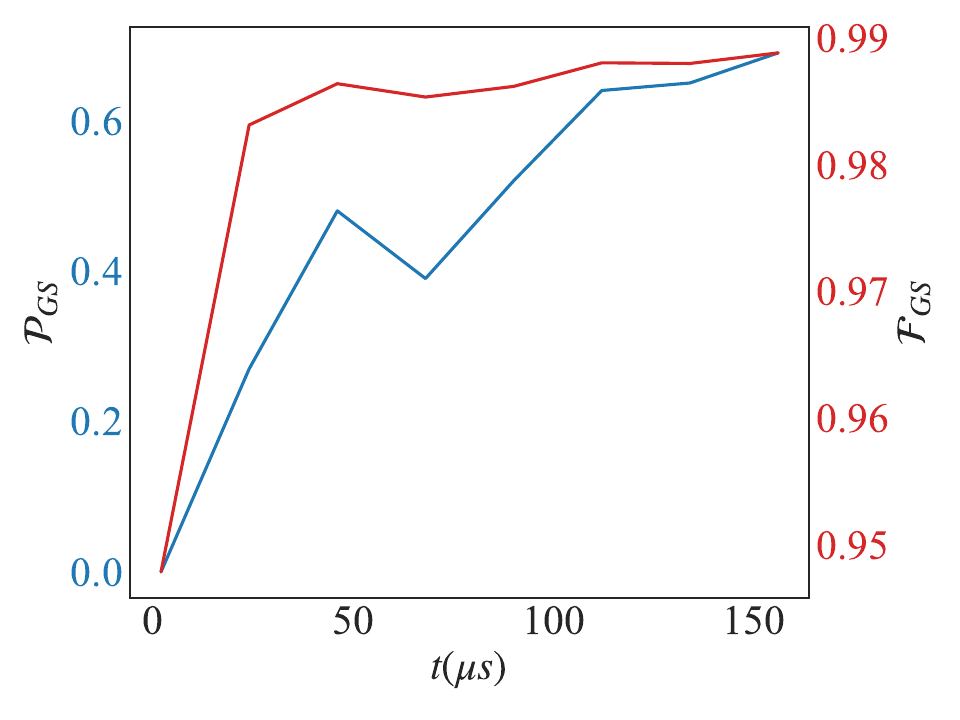}}%
    \subfloat[$h=0.5$]{\includegraphics[width=0.33\textwidth]{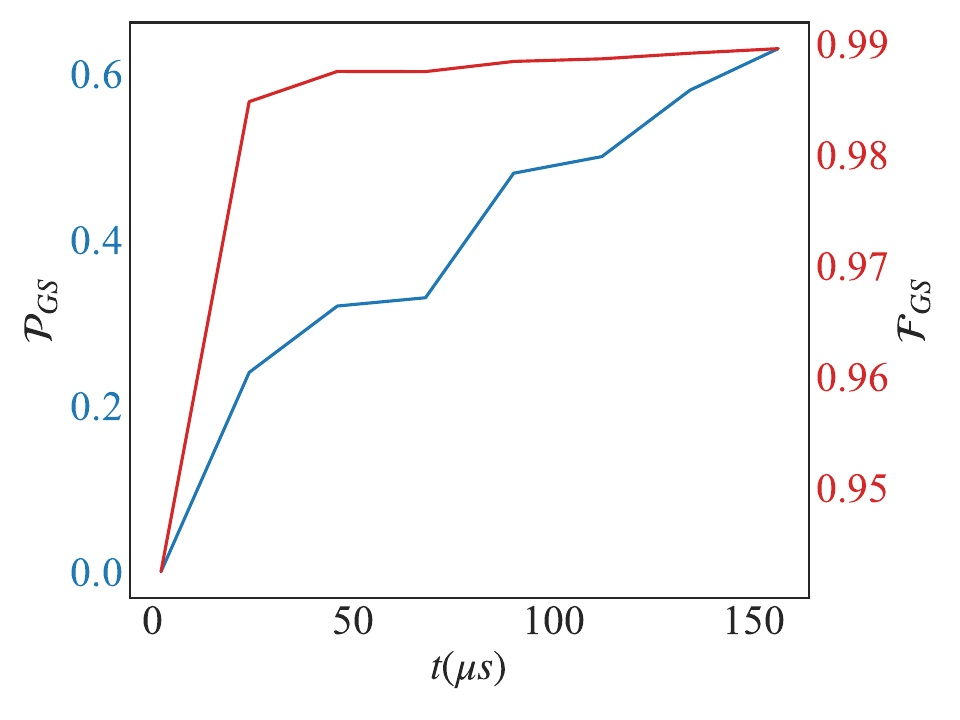}}%
    \subfloat[$h=1.0$]{\includegraphics[width=0.33\textwidth]{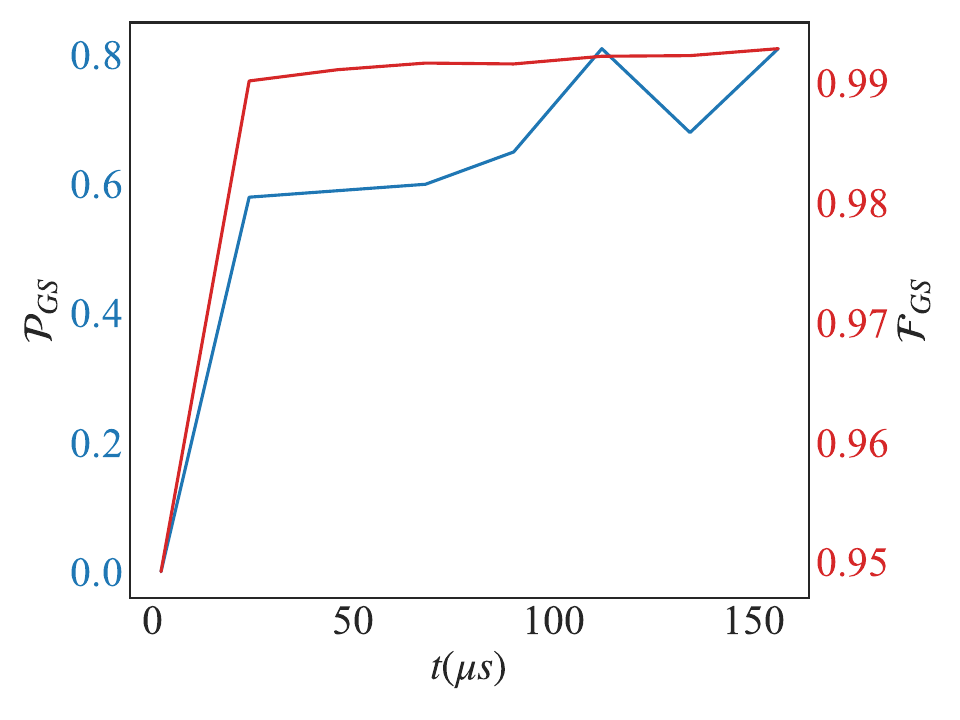}}
    \caption{The success probability $\mathcal{P}_{GS}$, Eq.~\eqref{pgs}, and the fidelity $\mathcal{F}_{GS}$, Eq.~\eqref{fgs}, under reverse-annealing without the pause for different values of the magnetic field $h$.  Each point is averaged over 100 annealing runs with 10 samples each.}
    \label{PF_reverse_h}
\end{figure*}
\begin{figure*}
    \centering
    \subfloat[$h=0.1$]{\includegraphics[width=0.33\textwidth]{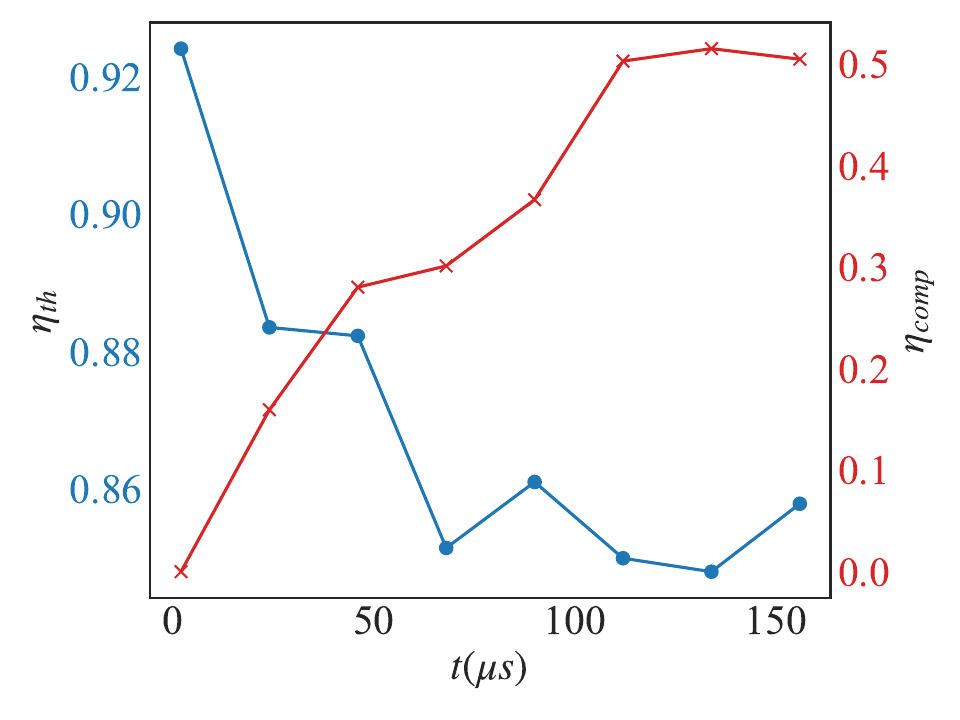}}
    \subfloat[$h=0.5$]{\includegraphics[width=0.33\textwidth]{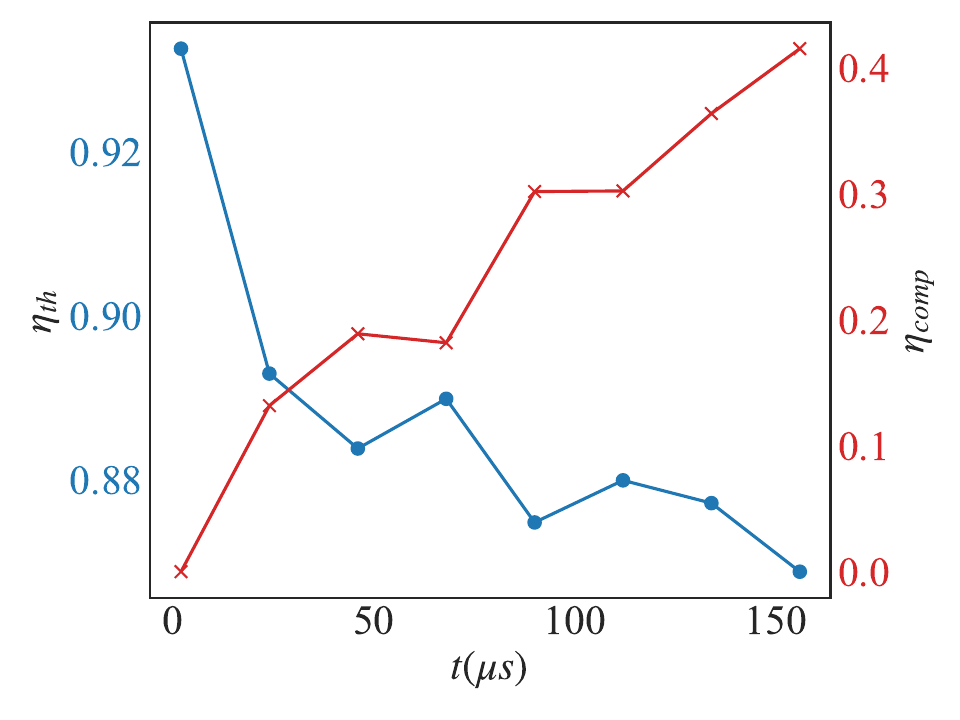}}%
    \subfloat[$h=1.0$]{\includegraphics[width=0.33\textwidth]{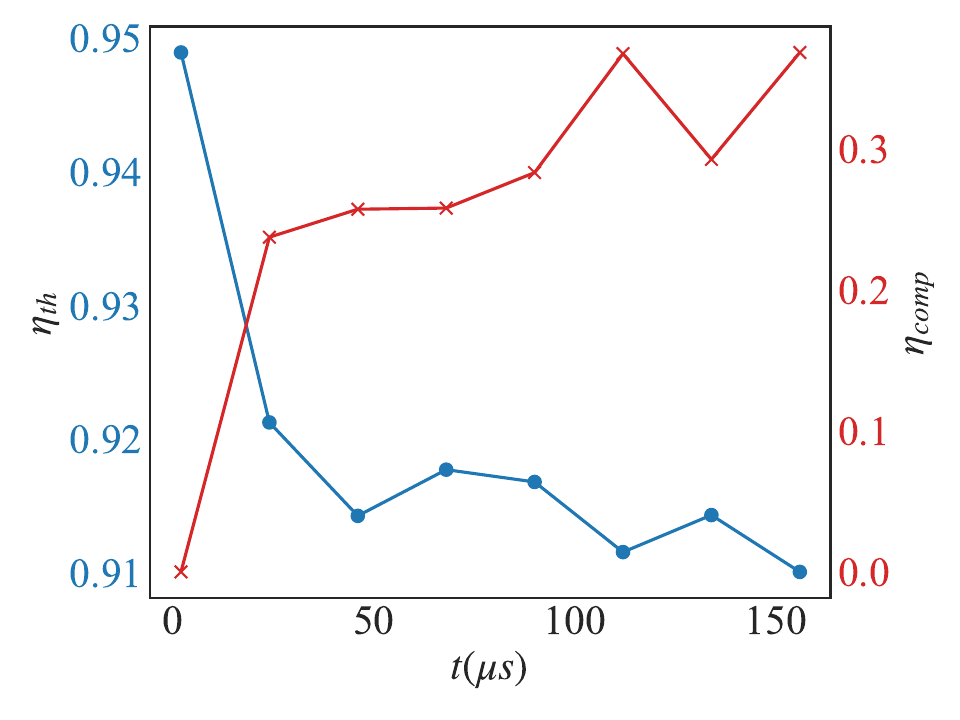}}
    \caption{The thermodynamic and computational efficiency, $\eta_{th}$ Eq.~\eqref{nth_eq} and $\eta_{comp}$ Eq.~\eqref{ncomp_eq} respectively, under reverse-annealing without the pause for different values of the magnetic field $h$.  Each point is averaged over 100 annealing runs with 10 samples each.}
    \label{nth_reverse_h}
\end{figure*}
\paragraph*{Reverse-annealing with pausing.}

The low computational performance of the chip for the simple Hamiltonian, Eq.~\eqref{ham}, shows that the reverse-annealing protocol does not exploit the energetics of the chip efficiently. Ideally, one aims at finding the protocol that provides a high computational efficiency at the lowest possible thermodynamical cost. For this reason, we introduce a pause in the reverse-annealing protocol, as depicted in Fig.~\eqref{QA_protocols}. Introducing a pause in the annealing schedule in quantum annealing has been shown to offer several benefits, such as: enhancing the probability of finding better solutions by efficient exploration of the solution space, which allows for a broader range of potential solutions to a given problem~\cite{pausing_lidar, pausing_rieffel}. Furthermore, since the pausing duration can be manipulated by the user, this offers the ability to balance between exploration and exploitation, which allow for the fine-tuning of the solution quality. The flexibility offered by the pausing strategy also allows for adaptation to the characteristics of specific problem instances, which enhances the efficiency and effectiveness of quantum annealing for a wide range of applications~\cite{pausing_2022}.

Figure~\eqref{reverse_no_h}(c+d) presents the results of applying a pause during the reverse-annealing schedule, as shown in Fig.~\eqref{QA_protocols}. The success probability $\mathcal{P_{GS}}$ improves dramatically as shown in panel (c), where it grows quickly to 0.8 during the annealing schedule, which means that 80\% of the 1000 annealing runs taken during the experiment returned the ground state energy. The fidelity, Eq.\eqref{fgs}, shown in the same panel also benefits from introducing the pause, where the overlap between the theoretical and experimental energy read is almost unity. The power of pausing is even more significant for the thermodynamic and computational efficiency, Eq.~\eqref{nth_eq} and Eq.~\eqref{ncomp_eq} respectively, reported in panel (d). We see that pausing allows for achieving high computational efficiency at a moderate thermodynamic cost, which is due to the concept of thermalization.  Introducing a pause in the annealing schedule allows the chip to  relax and thermalize after being excited by quantum or thermal effects near the minimum gap. However, pausing is not always beneficial, and it depends on several factors such as the relaxation rate, the pause duration, and the annealing schedule. The optimal protocol corresponds to a pause right after crossing the minimum gap and its duration should be no less than the thermalization time~\cite{pausing_rieffel}.

\subsection{Magnetically assisted annealing}

Next, we perform experiments with the magnetic field switched on, under reverse-annealing with and without pausing. The local magnetic field plays a crucial role in shaping the energy landscape and controlling the behavior of the qubits during the annealing process. By manipulating the local magnetic field, quantum annealers can explore and optimize complex problem spaces more effectively.

\paragraph*{Assisted reverse-annealing without pausing.} The benefit of introducing the magnetic field is clear from the behavior of the success probability $\mathcal{P}_{GS}$, Eq.~\eqref{pgs}, reported in Fig.~\eqref{PF_reverse_h}. In this case, without introducing a pause in the annealing schedule, the success probability of the ground state of the problem Hamiltonian is very high compared to the case when the magnetic field is off (c.f. Fig.~\eqref{reverse_no_h}(a)). Introducing the magnetic field influences the shape of the energy landscape that the qubits explore during the annealing process~\cite{watabe2020enhancing}. The landscape can be adjusted to promote or discourage certain configurations of the qubits, which can help guide the system toward the desired solution, and explains the slight improvement in the fidelity, Eq.~\eqref{fgs}, reported in the same panel. This dramatic improvement reflects itself on the thermodynamic and computational efficiency Eq.~\eqref{nth_eq} and Eq.~\eqref{ncomp_eq} respectively, of the chip as reported in Fig.~\eqref{nth_reverse_h}. In this case, introducing the magnetic field allows to guide the system in the energy landscape which is a more efficient strategy to exploit energy to perform computation.

\paragraph*{Assisted reverse-annealing with pausing.} 

\begin{figure*}
    \centering
    \subfloat[$h=0.1$]{\includegraphics[width=0.33\textwidth]{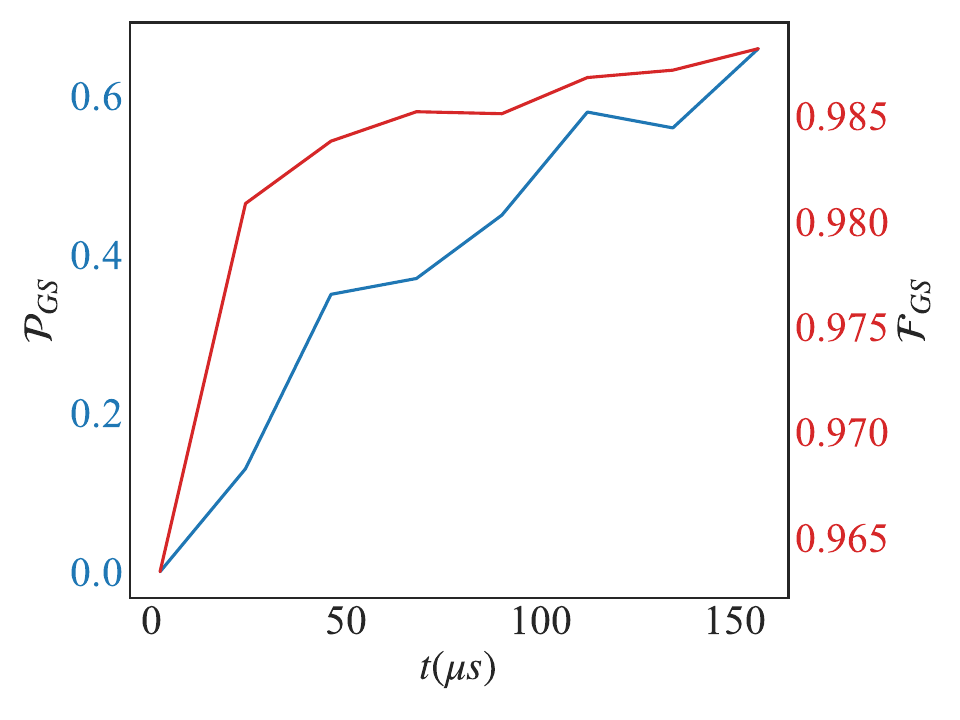}}%
    \subfloat[$h=0.5$]{\includegraphics[width=0.33\textwidth]{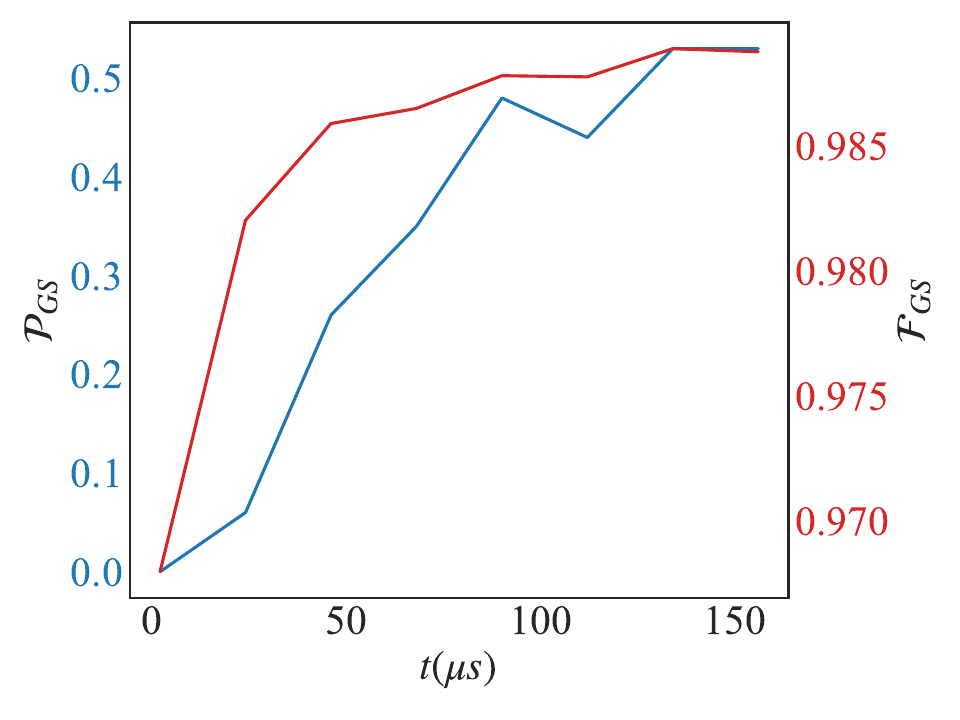}}%
    \subfloat[$h=1.0$]{\includegraphics[width=0.33\textwidth]{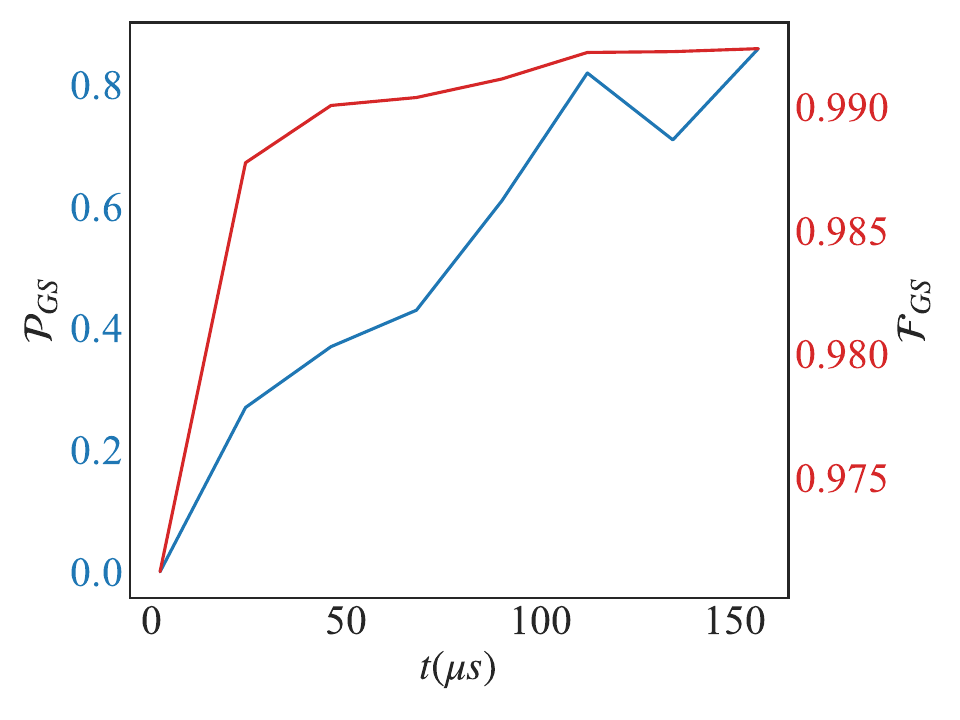}}
    \caption{The success probability $\mathcal{P}_{GS}$, Eq.~\eqref{pgs}, and the fidelity $\mathcal{F}_{GS}$, Eq.~\eqref{fgs},  under reverse-annealing with the pause for different values of the magnetic field $h$.  Each point is averaged over 100 annealing runs with 10 samples each.}
    \label{PF_pause_h}
\end{figure*}

\begin{figure*}
    \centering
    \subfloat[$h=0.1$]{\includegraphics[width=0.33\textwidth]{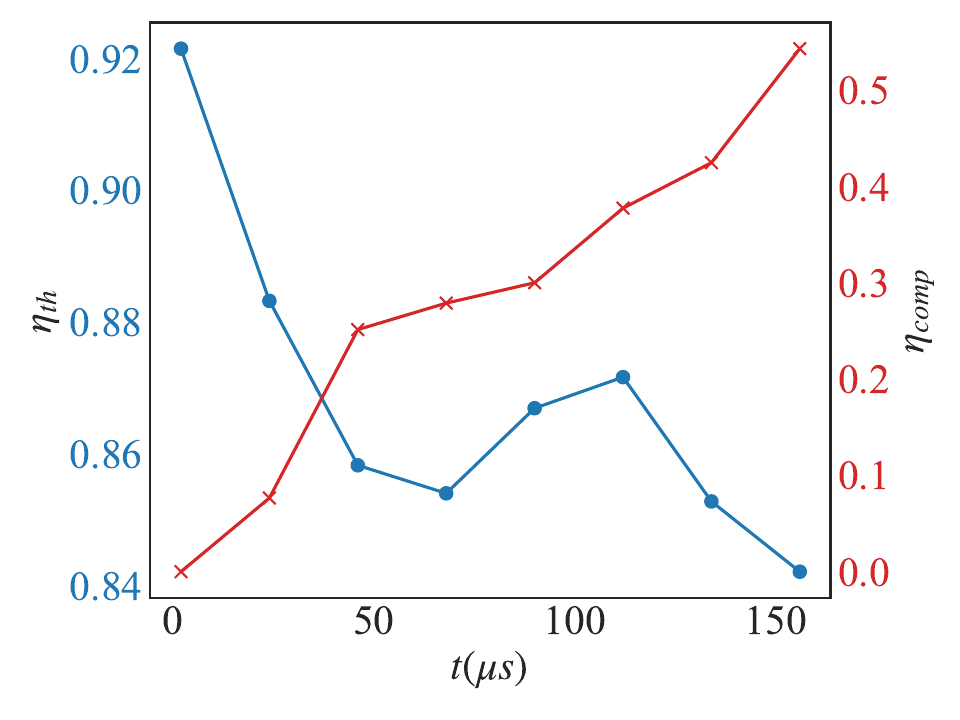}}
    \subfloat[$h=0.5$]{\includegraphics[width=0.33\textwidth]{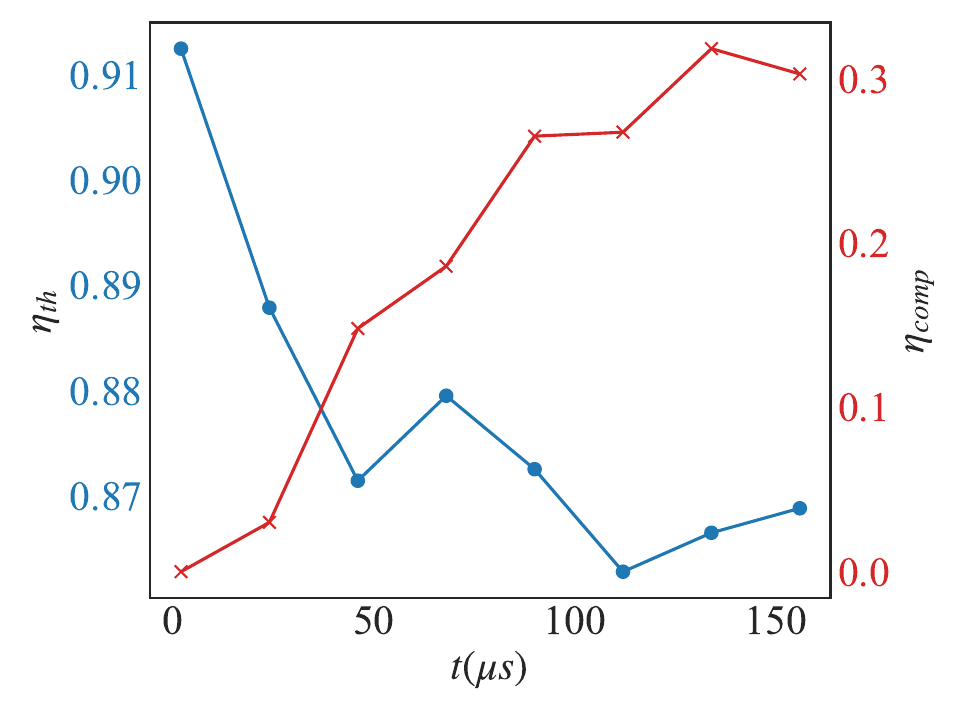}}%
    \subfloat[$h=1.0$]{\includegraphics[width=0.33\textwidth]{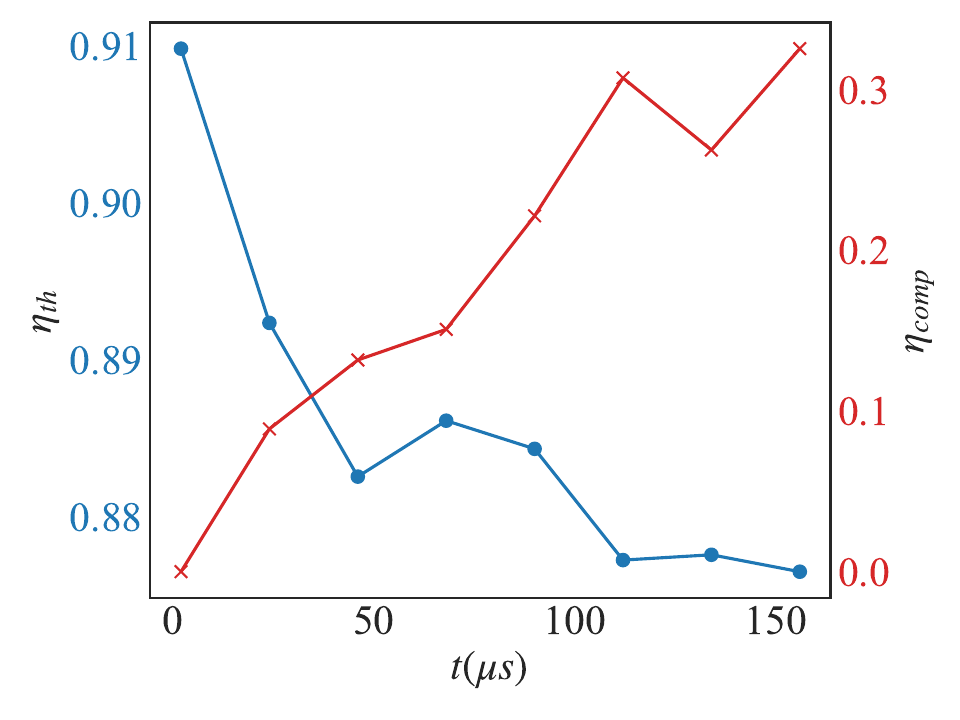}}
    \caption{The thermodynamic and computational efficiency, $\eta_{th}$ Eq.~\eqref{nth_eq} and $\eta_{comp}$ Eq.~\eqref{ncomp_eq} respectively, under reverse-annealing with the pause for different values of the magnetic field $h$.}
    \label{nth_pause_h}
\end{figure*}

Interestingly, in comparison with the case for $h=0$ reported in Fig.~\eqref{reverse_no_h}(c+d), introducing a pause in the annealing schedule with the magnetic field being present decreases the success probability $\mathcal{P}_{GS}$, Eq.~\eqref{pgs}, as can be seen from Fig.~\eqref{PF_pause_h}. Consequently, it decreases also the thermodynamic and computational efficiency, $\eta_{th}$ Eq.~\eqref{nth_eq} and $\eta_{comp}$ Eq.~\eqref{ncomp_eq} respectively, as can be seen from Fig.~\eqref{nth_pause_h}. Switching on the magnetic field in quantum annealing changes the qubit energy levels, and structure. On the other hand, for pausing to work it needs to be carefully applied while taking into account the energy level structure variation with the magnetic field. For this reason, the pause needs to be performed right after the minimum gap characterized by the value of the magnetic field $h$~\cite{pausing_lidar, pausing_rieffel}.

\section{\label{conclusion} Concluding remarks}

We have investigated the optimization of the computational efficiency and the thermodynamic cost in the D-Wave quantum annealing systems employing reverse-annealing. By combining reverse-annealing with pausing, we have demonstrated improved computational efficiency while operating at a lower thermodynamic cost compared to reverse-annealing alone. Our results highlight the potential benefits of strategically incorporating pausing into the annealing process to enhance overall computational and energetic performance. Furthermore, our results indicate that the magnetic field plays a crucial role in enhancing computational efficiency during reverse-annealing. However, when pausing is involved, the magnetic field becomes detrimental to the overall performance. This suggests the need for careful consideration of the magnetic field configuration and its impact on the energy gap of the system during the annealing process.

While our experiment was performed on the D-Wave Chimera architecture, it will be interesting to extend our experimental approach to the Pegasus and Zephyr architectures. These two models offer high tolerance to noise and a more complex structure, which allows us to investigate the trade-off between energetic performance and computational complexity. Additionally, exploring the scalability of these findings to larger-scale quantum systems and real-world applications remains a promising avenue for future research.

\acknowledgements{We would like to thank L. Buffoni and M. Campisi for valuable discussions. T.Ś. and Z.M. acknowledge support from the National Science Center (NCN), Poland, under Project No.~2020/38/E/ST3/00269. S.D. acknowledges support from  the U.S. National Science Foundation under Grant No. DMR-2010127 and the John Templeton Foundation under Grant No. 62422. B.G. acknowledges support from Foundation for Polish Science
(grant no POIR.04.04.00-00-14DE/ 18-00 carried out within the Team-Net program co-financed by the European Union under the European Regional Development Fund). The code of the experiments is publically available in the \href{https://github.com/tomsmierz/dwave-thermodynamics}{Github repository} \cite{git}.}

\bibliography{bib}

\end{document}